\newcommand{\MSbar}{\overline{\rm MS}}

\newcommand{\DRbarprime}{\overline{\rm DR}'}
\newcommand{\lnbar}{{\overline{\rm ln}}}
\newcommand\beq{\begin{eqnarray}}
\newcommand\eeq{\end{eqnarray}}
\newcommand{\myhat}{\widehat}
\newcommand{\gthree}{g_3}
\newcommand{\yt}{y_t}
\newcommand{\gthreehat}{\myhat g_3}
\newcommand{\ythat}{\myhat y_t}
\newcommand{\ghat}{\myhat g}
\newcommand{\gphat}{\myhat g'}
\newcommand{\mZhat}{\myhat m_Z}
\newcommand{\mWhat}{\myhat m_W}
\newcommand{\mthat}{\myhat m_t}
\newcommand{\mhhat}{\myhat m_h}
\newcommand{\vhat}{\myhat v}
\newcommand{\lambdahat}{\myhat\lambda}
\newcommand{\Bhht}{\myhat b_{h}}
\newcommand{\BhWt}{\myhat b_{W}}
\newcommand{\BhZt}{\myhat b_{Z}}
\newcommand{\bhht}{b_{h}}
\newcommand{\bhWt}{b_{W}}
\newcommand{\bhZt}{b_{Z}}
\newcommand{\Lt}{L}
\newcommand{\Lthat}{\myhat L}
\newcommand{\gluino}{m^2_{\tilde g}}
\newcommand{\sbotL}{m^2_{\tilde b_L}}
\newcommand{\sbeta}{s_{\beta}}
\newcommand{\cbeta}{c_{\beta}}
\newcommand{\sstop}{s_{\tilde t}}
\newcommand{\cstop}{c_{\tilde t}}
\newcommand{\stopone}{m^2_{\tilde t_1}}
\newcommand{\stoptwo}{m^2_{\tilde t_2}}
\newcommand{\fone}{f_1}
\newcommand{\ftwo}{f_2}
\newcommand{\clambda}{c_\lambda}
\newcommand{\cgA}{c_{g_3}}
\newcommand{\cyA}{c_{y_t}}
\newcommand{\cyB}{c'_{y_t}}
\newcommand{\cvev}{c_v}

\documentclass[twocolumn,
amsmath,
prd,nofootinbib,floatfix,
]{revtex4}

\allowdisplaybreaks
\usepackage{graphicx}

\begin{document}
\renewcommand{\theequation}{\arabic{section}.\arabic{equation}}

\title{Three-loop corrections to the lightest Higgs 
scalar boson mass in supersymmetry}

\author{Stephen P. Martin}
\affiliation{
Physics Department, Northern Illinois University, DeKalb IL 60115 USA\\
{\rm and}
Fermi National Accelerator Laboratory, PO Box 500, Batavia IL 60510}

\begin{abstract} 
I evaluate the largest three-loop corrections to the mass of the lightest 
Higgs scalar boson in the Minimal Supersymmetric Standard Model in a 
mass-independent renormalization scheme, using effective field theory and 
renormalization group methods.  The contributions found here are those 
that depend only on strong and Yukawa interactions and on the leading and 
next-to-leading logarithms of the ratio of a typical superpartner mass 
scale to the top quark mass.  The approximation assumes that all 
superpartners and the other Higgs bosons can be treated as much heavier 
than the top quark, but does not assume their degeneracy. I also discuss 
the consistent addition of the three-loop corrections to a complete 
two-loop calculation.
\end{abstract}

\maketitle

\tableofcontents

\section{Introduction}\label{sec:introduction}
\setcounter{equation}{0}
\setcounter{footnote}{1}

Low-energy supersymmetry breaking can stabilize the electroweak scale
against radiative corrections proportional to much higher mass scales,
including the Planck mass. In the minimal supersymmetric standard model
(MSSM), the lightest neutral Higgs scalar boson ($h$) mass is quartically
sensitive to the value of the top quark mass, but only logarithmically
sensitive to the scale of supersymmetry breaking, once the $Z$ boson mass
is taken as fixed. A future experimental determination of the masses and
couplings of the Higgs scalar bosons and the superpartners at the Fermilab
Tevatron $p\overline p$ collider, the CERN Large Hadron Collider and/or a
future linear $e^+ e^-$ collider will be crucial in understanding the
structure of supersymmetry breaking. 

The $h$ mass, in particular, is likely to be a very precisely measured
quantity \cite{HiggsLHC,HiggsLCa,HiggsLCe,HiggsLCj}. This has motivated
many studies of the relationship between the physical Higgs mass $M_{h}$
and the underlying Lagrangian parameters, in the form of radiative
corrections of increasing precision and detail
\cite{Li:1984tc}-\cite{Frank:2006yh}.  The tremendous effort that has been
expended on these calculations is necessitated by the appearance of
qualitatively new enhancement effects at each of the first two loop orders
in perturbation theory.  The tree-level result depends only on electroweak
gauge couplings, which enter into the quartic Higgs coupling. At one-loop
order, the large top Yukawa coupling, enhanced by a color factor, enters.
At two-loop order, the QCD coupling makes its appearance. It turns out
that even three-loop order contributions will be necessary if the goal is
to make purely theoretical errors negligible compared to the future
experimental uncertainty in $M_{h}$. (Of course, there will also be
important sources of error due to a lack of precise knowledge of the input
parameters of the theory, such as the top-quark Yukawa couplings and the
soft supersymmetry breaking terms in the Lagrangian; these are considered
as experimental errors for the present discussion.)

Three general methods have been commonly used, often in combination, for
evaluating $M_{h}$. First, the pole mass can be computed by a
straightforward calculation of the neutral Higgs self-energy diagrams. The
resulting complete expressions are quite complicated and unwieldy beyond
one-loop order. A second approach uses the effective potential
approximation. This means that radiative corrections to $M^2_{h}$ are
computed by taking the second derivatives of the effective potential; this
is equivalent to computing the pole mass from self-energy functions in the
approximation that the external momentum is neglected. This has the
advantage that calculations can be reduced to vacuum graphs, which can
always be analytically computed through 2-loop order. However, this method
is not gauge-fixing invariant, has limited accuracy, and can suffer from
numerical instabilities if one chooses a renormalization scale at which a
tree-level squared mass happens to be extremely small. A third method uses
the method of effective Lagrangians, with renormalization group running
used to systematically isolate the effects that are enhanced by logarithms
of ratios of the superpartner mass scale to the electroweak and top-quark
mass scales. Recent reviews and descriptions of computer programs
implementing some of the known results can be found in
\cite{FeynHiggs,Lee:2003nt,Allanach:2004rh,%
Heinemeyer:2004ms,Frank:2006yh}.

In this paper, I will use a combination of the three methods mentioned
above to evaluate the most important 3-loop contributions to $M_{h}$. The
input parameters in this result will be the running $\DRbarprime$
\cite{DRbar,DRbarprime} parameters in the full theory with no
superpartners decoupled. (These results can also be converted into
on-shell or hybrid schemes in which all or some of the input particle
masses are taken to be physical masses rather than running masses,
although that is not done explicitly here.) The results can be used to
supplement my previous evaluation of $M_{h}$ at two-loop order, which
includes the full diagrammatic results that involve the strong
interactions and the Yukawa interactions (including the ones that also
involve the electroweak couplings) \cite{SPM0405022}, together with all
other two-loop contributions in the effective potential approximation
\cite{SPM0206136,SPM0211366}. 

The 3-loop contributions to be found here are only the ones that are
proportional to powers of the strong coupling and the top-quark Yukawa
coupling. Also, only the contributions containing the leading and
next-to-leading powers of $\ln( Q^2_{\rm SUSY}/m_{t}^2)$ at 3-loop order
are evaluated, where $m_t$ is the top-quark mass and $Q_{\rm SUSY}$ is a
renormalization scale comparable to a typical superpartner mass scale.
However, I will not assume that the superpartners are degenerate or that
top-squark mixing is negligible. These logarithmically enhanced
contributions are likely to dominate, numerically. While there is no
unsurmountable obstacle to evaluating the analogous contributions at
arbitrary loop order using the same methods, as a practical matter they
are unlikely to be as large as the remaining uncalculated 3-loop
corrections, nor are they likely to be as large as the practical
experimental errors once input parameter uncertainties are taken into
account.

\section{Conventions and setup} 
\setcounter{equation}{0} 
\setcounter{footnote}{1}

In the following,
\beq
\kappa &\equiv& 1/16 \pi^2
\eeq
is used as a loop factor, and the renormalization scale is denoted
$Q$. Also, I define the symbol
\beq
\lnbar(x) =\ln (x/Q^2).
\eeq 
Some formulas below will make use of the one-loop vacuum integral
function
\beq
A(x) = x \lnbar (x) - x,
\eeq
and the Passarino-Veltman one-loop self-energy scalar loop integral with
external momentum invariant $s$ and equal internal squared masses $x$,
\begin{widetext}
\beq
f_B(s,x,q^2) &=& 
\left \{ \begin{array}{ll}
 2 - \ln (x/q^2) - 2 (4x/s -1)^{1/2} \sin^{-1}(\sqrt{s/4x})
 & (s \leq 4 x)
\\[6pt]
 2 - \ln (x/q^2) + (1 - 4x/s)^{1/2}
 \bigl \{ \ln (s[1 - (1 - 4x/s)^{1/2}]/2x -1) + i \pi \bigr \}
\phantom{xxxxx}
 & (s > 4x) .
\end{array}
\right.
\eeq
\end{widetext}
(This function will appear in some formulas below with $q$ not equal to
the renormalization scale $Q$.) This has the small $s$ expansion, valid
for $s<4x$: 
\beq
f_B(s,x,q^2) &=& -\ln(x/q^2) + s/(6x) + s^2/(60 x^2) 
\phantom{xxx}
\nonumber \\ &&
+ s^3/(420 x^3) + \ldots ,
\label{eq:expfB}
\eeq
and the special value:
\beq
f_B(x,x,q^2) &=& 2 - \pi/\sqrt{3} - \ln (x/q^2) .
\eeq

The neutral Higgs complex scalar field $\phi$ of the Standard Model
effective theory has a tree-level potential
\beq
V = -\myhat m^2_\phi |\phi|^2 + \frac{\lambdahat}{4} |\phi|^4.
\eeq
where $\myhat m_\phi$ and $\lambdahat$ are running $\MSbar$ parameters. 
The top quark and gauge interactions of the effective Standard Model
theory are governed by $\MSbar$ running parameters: 
\beq
\ythat, \>\>\gthreehat, \>\>\ghat, \>\>\gphat, \>\> \vhat,
\eeq
with other Yukawa couplings neglected. The minimum of the tree-level
potential occurs at $\langle \phi \rangle = 2 \myhat m^2/\lambdahat$.
However, here I expand instead around the vacuum expectation value (VEV)
$\vhat$ defined as the minimum of the loop-corrected Landau gauge
effective potential of the theory. Explicitly,
\beq
\lambdahat \vhat^2 &=& 2 \myhat m^2_\phi 
- \frac{2}{\vhat} \frac{\partial}{\partial\phi}
\Delta V_{\rm eff} (\phi,\phi^*)
\Bigl |_{\phi=\phi^*=\vhat}\Bigr. .
\label{eq:defvhat}
\eeq
For example, working at one-loop order, 
\beq
\lambdahat \vhat^2 &=& 2 \myhat m^2_\phi + \kappa
\Bigl \lbrace 12 \ythat^2 A(\mthat^2) 
- \frac{3\lambdahat}{2} A(\mhhat^2)
\nonumber \\ &&
-\frac{1}{2}(\ghat^2 + \gphat^2)[3 A(\mZhat^2) +2 \mZhat^2]
\nonumber \\ &&
-\ghat^2 [3 A(\mWhat^2) +2 \mWhat^2]
\Bigr \rbrace + {\cal O}(\kappa^2),
\label{eq:oneloopdefvhat}
\eeq
where
\beq
\myhat m^2_{h} &=& \lambdahat \vhat^2,\\
\mthat &=& \ythat \vhat \\
\mWhat^2 &=& \ghat^2 \vhat^2/2\\
\mZhat^2 &=& (\ghat^2 + \gphat^2) \vhat^2/2.
\eeq
Equation (\ref{eq:defvhat}) is used to eliminate $\myhat m^2_\phi$ in
favor of $\vhat$.  The normalization of the Higgs VEV is such that $\vhat$
is roughly 175 GeV. 

The MSSM theory is governed by (unhatted) running $\DRbarprime$ parameters
including the gauge couplings, top-quark Yukawa coupling, and Higgs
expectation values (defined as the minimum of the loop-corrected Landau
gauge effective potential of the full MSSM theory): 
\beq
 \gthree,\>\> g,\>\> g',\>\> y_t,\>\> v_u,\>\> v_d,
\label{eq:MSSMparams}
\eeq
The last three of these are taken to be real and positive by convention. 
The parameters 
\beq
m_t &\equiv& y_t v_u\\
v &\equiv& (v_u^2 + v_d^2)^{1/2}\\
\tan(\beta) &\equiv& v_u/v_d
\eeq
are defined in terms of them, and so depend on the renormalization scale
$Q$. In the following, I will also use the short-hand notations
\beq
s_\beta = \sin(\beta),
\qquad
c_\beta = \cos(\beta),
\qquad
c_{2\beta} = \cos(2\beta).\phantom{xx}
\eeq
The top-squark sector has a tree-level running squared-mass matrix
in the $(\tilde t_L, \tilde t_R)$ basis:
\beq
\begin{pmatrix}
m^2_{\tilde t_L} + m_t^2 \phantom{xx}& v_u a_t^* - v_d \mu y_t 
\cr
v_u a_t - v_d \mu^* y_t \phantom{xx} & m^2_{\tilde t_R} + m_t^2
\end{pmatrix}
\eeq
where electroweak $D$-terms are neglected (appropriately for the 
approximation used below), and the notation follows 
\cite{SPM0206136,primer}.
The mass eigenstates are related to the gauge eigenstates by
\beq
\begin{pmatrix}
\tilde t_L \cr \tilde t_R
\end{pmatrix}
&=& 
\begin{pmatrix}
c_{\tilde t} \phantom{x}& -s_{\tilde t}^* 
\cr
s_{\tilde t} \phantom{x} & c_{\tilde t}^*
\end{pmatrix}
\begin{pmatrix}
\tilde t_1 \cr \tilde t_2
\end{pmatrix},
\eeq
with $|c_{\tilde t}|^2 + |s_{\tilde t}|^2 = 1$.  (Here I use the
conventions of ref.~\cite{SPM0206136}; those of ref.~\cite{primer} are
related by $s_{\tilde t} \rightarrow -s_{\tilde t}$.) If $\mu$ and $a_t$
are real, then $s_{\tilde t}$ and $c_{\tilde t}$ are the sine and cosine
of a top-squark mixing angle; otherwise they can be complex (but
$c_{\tilde t}$ can always be taken real as a convention). It is convenient
to define the parameter $X_t$ by: 
\beq
m_t X_t = v_u a_t - v_d \mu^* y_t 
= -s_{\tilde t} c_{\tilde t}^* (m^2_{\tilde t_2} - m^2_{\tilde t_1}),
\eeq
in terms of which the squared-mass eigenvalues are:
\beq
m^2_{\tilde t_1}, m^2_{\tilde t_2} 
&=& \frac{1}{2} \bigl [
m^2_{\tilde t_R} + m^2_{\tilde t_L} + 2 m_t^2 
\nonumber \\ &&
\mp
\lbrace
(m^2_{\tilde t_R} - m^2_{\tilde t_L})^2 + 4 m_t^2 |X_t|^2 \rbrace^{1/2} \bigr ]
.
\phantom{xxxx.}
\eeq
Also appearing below are the tree-level running squared-mass eigenvalues 
of the gluino and squarks, denoted 
\beq
m_{\tilde g}^2,\>\>\> m^2_{\tilde q_i}\quad(i=1,\ldots,12).
\eeq
The latter include the squared mass eigenvalues of $\tilde t_1$ and
$\tilde t_2$, as well as the other squarks which are taken to be unmixed.
The neutralino and chargino mass eigenstates will also be taken to be
unmixed, with the Higgsinos having a common squared mass $|\mu|^2$. The
electroweak gauginos do not contribute in the approximation used here.
Likewise, the Higgs scalar bosons $H^0, A^0, H^\pm$ are treated in the
decoupling limit, with a common running squared mass $m_H^2$ (supposed to
be much larger than $m_{h}^2$ and $m_t^2$), and mixing angle $\alpha =
\beta - \pi/2$. 


\section{Higgs pole mass in the Standard Model\label{sec:SM}} 
\setcounter{equation}{0} 
\setcounter{footnote}{1}

To prepare for matching the Standard Model to the MSSM, one can use the
renormalization group to obtain the higher-loop contributions of leading
and next-to-leading order in
\beq
\Lthat \equiv \ln(Q^2/\mthat^2)
\eeq
for large $Q$.  This is done by using the fact that $M^2_h$ is an
observable and therefore renormalization-scale independent. Let us write: 
\beq
M^2_h = \sum_{n=0}^\infty \kappa^n \sum_{p=0}^n \Lthat^p C_{n,p}
\eeq
where the quantities $C_{n,p}$ only depend on $Q$ implicitly through the
running parameters.  For $p\not=0$, the coefficient $C_{n,p}$ can be
obtained from the results with smaller $n$, provided that the $n-p+1$ loop
order beta functions for each of the running parameters $X$ are known. The
3-loop order beta functions for the scalar sector of the Standard Model
are evidently not available at present, so only the leading and
next-to-leading contributions in $\Lthat$ can be found at each loop order
in this way. In general, they satisfy recursion relations: 
\beq
C_{n,n} &=& -\frac{1}{2n} \sum_X \beta_X^{(1)}
\frac{\partial}{\partial X} C_{n-1,n-1},
\label{eq:Cnn}
\eeq
\begin{widetext}
\beq
C_{n,n-1} &=& \bigl[
\beta_{\vhat}^{(1)}/\vhat + \beta^{(1)}_{\ythat}/\ythat \bigr ] 
C_{n-1,n-1} 
-
\frac{1}{2(n-1)} \sum_X \Bigl [
\beta_X^{(1)} \frac{\partial}{\partial X} C_{n-1,n-2}
+ 
\beta_X^{(2)} \frac{\partial}{\partial X} C_{n-2,n-2}
\Bigr ],
\eeq
\end{widetext}
where
\beq
\beta_{X} \,\equiv\, Q \frac{dX}{dQ} \,=\, 
\kappa \beta^{(1)}_X
+ \kappa^2 \beta^{(2)}_X + \ldots
\eeq
for $X= \ythat, \gthreehat, \vhat, \lambdahat, \ghat, \gphat$. In the
following, dependence on $\ghat, \gphat$ will be dropped, so the pertinent
2-loop renormalization group equations for the Standard Model parameters
are \cite{Machacek:1984zw,FJJ}: 
\beq
\beta^{(1)}_{\lambdahat} &=& 
-24 \ythat^4 + 12 \lambdahat \ythat^2 + 6 \lambdahat^2 
,
\\
\beta^{(2)}_{\lambdahat} &=&
-128 \gthreehat^2 \ythat^4 
+ 120 \ythat^6 
+ 80 \gthreehat^2 \ythat^2 \lambdahat
\nonumber \\ &&
- 3 \ythat^4 \lambdahat 
- 36 \ythat^2 \lambdahat^2
- 39 \lambdahat^3/2 
,
\\
\beta^{(1)}_{\ythat} &=&
9\ythat^3/2 - 8 \gthreehat^2 \ythat 
,
\\
\beta^{(2)}_{\ythat} &=&
-108 \gthreehat^4 \ythat 
+ 36 \gthreehat^2 \ythat^3 
-12 \ythat^5 
\nonumber \\ &&
-3 \ythat^3 \lambdahat 
+ 3 \ythat \lambdahat^2/8 
,
\\
\beta^{(1)}_{\vhat} &=& 
-3 \ythat^2 \vhat 
,
\\
\beta^{(2)}_{\vhat} &=&
\bigl (-20 \gthreehat^2 \ythat^2 + 27 \ythat^4/4 - 3 \lambdahat^2/8
     \bigr ) \vhat
,
\\
\beta^{(1)}_{\gthreehat} &=& -7\gthreehat^3
,
\\ 
\beta^{(2)}_{\gthreehat} &=& -26 \gthreehat^5 - 2 \gthreehat^3 \ythat^2
.
\label{eq:betagthreehattwo}
\eeq

In the Standard Model, a routine calculation shows that the one-loop pole
squared mass of the Higgs boson in the $\MSbar$ scheme can be written as: 
\begin{widetext}
\beq
M_{h}^2 &=& \lambdahat \vhat^2 + \kappa  \Bigl \lbrace
3 \ythat^2 (4 \mthat^2 - \mhhat^2) f_B(\mhhat^2,\mthat^2,Q^2)
- \frac{9}{4} \lambdahat \mhhat^2 f_B(\mhhat^2,\mhhat^2,Q^2)
\nonumber \\ &&
+ \frac{1}{2}[\lambdahat (4\mWhat^2 - \mhhat^2) 
-6 \ghat^2 \mWhat^2 ] f_B(\mhhat^2,\mWhat^2,Q^2)
+ \frac{1}{4}[\lambdahat (4 \mZhat^2 - \mhhat^2)
-6 (\ghat^2 + \gphat^2) \mZhat^2 ] f_B(\mhhat^2,\mZhat^2, Q^2)
\nonumber \\ &&
-\lambdahat [A(\mWhat^2) + A(\mZhat^2)/2] 
+ 2 \ghat^2 \mWhat^2 + (\ghat^2 + \gphat^2) \mZhat^2 
\Bigr \rbrace .\phantom{xxx}
\label{eq:SMMh}
\eeq
\end{widetext}
Then, writing $C_{n,p} = c_{n,p} \vhat^2$, one can choose: 
\beq
c_{0,0} &=& \lambdahat,
\label{eq:czerozeroSM}
\\
c_{1,0} &=& 2 \lambdahat \ythat^2 - \lambdahat^2
(9 \Bhht + 2 \BhWt + \BhZt + 6/5) /4
\phantom{xxxx}
\nonumber \\ &&
+ {\cal O}(\lambdahat^3)
,
\eeq
from which it follows, via 
eqs.~(\ref{eq:Cnn})-(\ref{eq:betagthreehattwo}), that:
\beq
c_{1,1} &=& 
12 \ythat^4 - 3 \ythat^2 \lambdahat - 3 \lambdahat^2
,
\\
c_{2,2} &=& 96 \gthreehat^2\ythat^4  - 54 \ythat^6 
- \lambdahat (12 \gthreehat^2 \ythat^2  + 99 \ythat^4/4) 
\nonumber \\ &&
+ 18 \lambdahat^2 \ythat^2 + 9 \lambdahat^3
,
\phantom{xxx}
\\
c_{2,1} &=& -32 \gthreehat^2 \ythat^4  - 18 \ythat^6 
+  \lambdahat \bigl (20 \gthreehat^2 \ythat^2
+ \ythat^4 [81/20 
\nonumber \\ &&
-54 \Bhht - 12 \BhWt - 6 \BhZt ] \bigr )
+ {\cal O}(\lambdahat^2),
\\
c_{3,3} &=& 
736 \gthreehat^4 \ythat^4 - 672 \gthreehat^2 \ythat^6+ 90 \ythat^8 
+ \lambdahat (-60 \gthreehat^4 \ythat^2 
\phantom{xxxx}
\nonumber \\ &&
- 102 \gthreehat^2 \ythat^4
+ 243 \ythat^6) 
+ {\cal O}(\lambdahat^2)
,
\\
c_{3,2} &=& 160 \gthreehat^4 \ythat^4 + 168 \gthreehat^2 \ythat^6
+ [-324 \Bhht -72 \BhWt 
\nonumber \\ &&
-36 \BhZt + 6633/10] \ythat^8 
+ {\cal O}(\lambdahat),
\\
c_{4,4} &=& 
5520 \gthreehat^6 \ythat^4 - 6492 \gthreehat^4 \ythat^6 
+ 2178 \gthreehat^2 \ythat^8
\nonumber \\ && + 783 \ythat^{10}/2 
+ {\cal O}(\lambdahat) .
\label{eq:cfourfourSM}
\eeq
Here, dependences on $\ghat,\gphat$ have been dropped
except where they enter through the kinematic quantities
\beq
\Bhht &=& f_B(\mhhat^2,\mhhat^2,\mthat^2),
\\
\BhWt &=& f_B(\mhhat^2,\mWhat^2,\mthat^2),
\\
\BhZt &=& f_B(\mhhat^2,\mZhat^2,\mthat^2),
\label{eq:BhZt}
\eeq
and $f_B(\mhhat^2,\mthat^2,\mthat^2)$ has been expanded using
eq.~(\ref{eq:expfB}). Note that terms up to order $\lambdahat^{N-n+k}$ in
the coefficients $c_{n,n}$ and $c_{n,n-1}$ are needed to generate the
terms of order $\lambdahat^k$ in the coefficients $c_{N,N}$ and
$c_{N,N-1}$ for $N>n$. However, only terms of order $\lambdahat$ in
$C_{n,n}$ and independent of $\lambdahat$ in $C_{n,n-1}$ will be needed in
the next section, because $\lambdahat$ is proportional to electroweak
couplings at tree level in the MSSM. The running parameters in
eqs.~(\ref{eq:czerozeroSM})-(\ref{eq:BhZt}) are all evaluated at the same
arbitrary renormalization scale $Q$ appearing in $L$. 

It should be emphasized that the expansion given above for $M_h^2$ is far
from unique. In particular, the seed expression for $c_{1,0}$ could have
been chosen differently, corresponding e.g.~to trading running masses for
physical masses in the one-loop correction part of eq.~(\ref{eq:SMMh}).
This would produce a different set of higher coefficients, but the
expression for the total physical mass $M^2_h$ would differ only by an
amount consistently neglected within the approximations. Here, I have
chosen to do the expansion entirely in terms of running parameters. 

\section{Matching to the MSSM\label{sec:matching}} 
\setcounter{equation}{0} 
\setcounter{footnote}{1}

The result for the Higgs pole squared mass $M_h^2$ in the previous section
can now be used to obtain an approximate formula in terms of the running
$\DRbarprime$ parameters of the MSSM. To do this, one needs the one-loop
matching conditions, which can be written in the form: 
\beq
\lambdahat &=& 
\frac{1}{2} (g^2 + g^{\prime 2}) c_{2\beta}^2 
+ \kappa \yt^4 \sbeta^4 \clambda 
+ \ldots
,
\\
\vhat &=& (v_u/\sbeta) \left [
1 + \kappa \yt^2 \sbeta^2 \cvev + \ldots 
\right ] ,
\\
\ythat &=& \yt \sbeta 
\left [1 + \kappa \left (\gthree^2 \cyA  + \yt^2 \sbeta^2 \cyB \right ) + 
\ldots
\right ]
,
\\
\gthreehat &=& \gthree \left [1 + \kappa \gthree^2\cgA  + \ldots \right ]
 .
\eeq
The matching coefficients $\clambda$, $\cvev$, $\cyA$, $\cyB$, and $\cgA$
depend on the renormalization scale $Q$, which is arbitrary but is taken
to be comparable to the superpartner and heavy Higgs bosons ($A^0$, $H^0$,
$H^\pm$) masses. All of these scales are assumed to be much larger than
the top and lightest Higgs ($h$) and electroweak gauge boson masses, so
that effects suppressed by powers of $m_{\tilde t_{1,2}}$, $m_{\tilde g}$,
$m_H$, $|\mu|$, etc., are neglected.  Then one can work consistently to
next-to-leading order in
\beq
\Lt = \ln(Q^2/m_t^2),
\eeq
that is, keeping $\Lt^n$ and $\Lt^{n-1}$ in the terms of $n$ loop order.
Eliminating the Standard Model parameters in favor of MSSM 
parameters, one obtains:
\beq 
M_h^2 = m_h^2 
+ m_t^2 y_t^2 s_\beta^2 \sum_{n=1}^\infty \kappa^n \Delta_{n}
,
\label{eq:M2hMSSM}
\eeq
with
\begin{widetext}
\beq
m_h^2 &=& \frac{1}{2} (g^2 + g^{\prime 2}) c_{2\beta}^2 v^2 ,
\label{eq:M2hzeroX}
\\
\Delta_{1} &=& 
12 \Lt + \clambda 
,
\label{eq:M2honeX}
\\
\Delta_{2} &=& 
( 96 g_3^2 - 54 y_t^2 \sbeta^2 ) \Lt^2
+\bigl [ (48 \cyA - 32) g_3^2 
+ (48 \cyB + 24 \cvev - 3 \clambda -18 ) y_t^2 \sbeta^2 \bigr ]\Lt
+ \ldots
,
\label{eq:M2htwoX}
\\
\Delta_{3} &=& 
( 736 g_3^4 - 672 g_3^2 y_t^2 \sbeta^2 + 90 y_t^4 \sbeta^4) \Lt^3
+\bigl [ 
  (160 + 192 \cgA + 384 \cyA) g_3^4
\nonumber \\ &&
 + ( 168 -12 \clambda - 324 \cyA +384 \cyB + 192 \cvev ) g_3^2 y_t^2 \sbeta^2 
\nonumber \\ &&
 + (6633/10 - 324 \bhht - 72 \bhWt - 36 \bhZt - 99\clambda/4 -324 \cyB 
   -108 \cvev\bigr )y_t^4 \sbeta^4 
\bigr ]\Lt^2 + \ldots ,
\label{eq:M2hthreeX}
\\
\Delta_{4} &=& 
(5520 g_3^6  - 6492 g_3^4 y_t^2 \sbeta^2 
+ 2178 g_3^2 y_t^4\sbeta^4 
+ 783 y_t^{6} \sbeta^6 /2)L^4 + \ldots, 
\label{eq:M2hfourX}
\eeq
where dependences on $g$ and $g'$ have been dropped in the loop
corrections, except where they enter through the kinematic quantities
\beq
\bhht &=& f_B(m_h^2,m_h^2,m_t^2),
\\
\bhWt &=& f_B(m_h^2,m_W^2,m_t^2),
\\
\bhZt &=& f_B(m_h^2,m_Z^2,m_t^2).
\eeq
(For later comparison purposes, the leading-logarithm four loop
contribution is also included.) It is important to note that the validity
of the result just given requires that the expansion is made in terms of
running couplings and masses, always evaluated at the renormalization
scale $Q$. Indeed, this requirement even applies to the terms that are not
written here explicitly because they are suppressed by electroweak gauge
couplings. [For example, there are terms at one-loop order proportional to
electroweak couplings multiplied by $f_B(m_h^2,m_Z^2,m_t^2)$. One could
re-express those contributions in terms of, for example,
$f_B(M_h^2,M_Z^2,M_t^2)$ involving the physical masses $M_h$, $M_Z$ and
$M_t$, but that would require changing the $\kappa^3 L^2$ coefficient
appearing in the expansion above.]

It remains to find the matching coefficients appearing in the above
expressions. First, $\clambda$ can be evaluated by comparing the
well-known one-loop Higgs pole mass calculated directly in the MSSM to
eqs.~(\ref{eq:M2hMSSM})-(\ref{eq:M2honeX}), giving: 
\beq
\clambda &=& 6 
\Bigl [\lnbar (m^2_{\tilde t_1}) + \lnbar (m^2_{\tilde t_2})
+ 2|X_t|^2 \ln(m^2_{\tilde t_2}/m^2_{\tilde t_1})
/( m^2_{\tilde t_2} -  m^2_{\tilde t_1}) 
\nonumber \\ &&
+ |X_t|^4 \bigl \lbrace 2 -
[(m^2_{\tilde t_2} +  m^2_{\tilde t_1})/(m^2_{\tilde t_2} - m^2_{\tilde t_1})] 
   \ln(m^2_{\tilde t_2}/m^2_{\tilde t_1}) 
 \bigr \rbrace /(m^2_{\tilde t_2} - m^2_{\tilde t_1})^2
\Bigr ] 
.
\eeq
The coefficients $\cyA$ and $\cyB + \cvev$ are obtained by equating the
one-loop expressions for the top-quark pole mass as computed in the full
MSSM and in the effective Standard Model theory, with the result: 
\beq
\cyA &=& \frac{4}{3} \Bigl [1 
+ 2 \fone (\gluino, \stopone)
+ 2 \fone (\gluino, \stoptwo)
+ 2 {\rm Re}[X_t] m_{\tilde g} 
\lbrace
\ftwo (\gluino, \stoptwo) - \ftwo (\gluino, \stopone)
\rbrace /(\stoptwo - \stopone)
\Bigr ] 
,
\phantom{xxx}
\\
\cyB + \cvev &=&  \Bigl [ \frac{3}{4} \cbeta^2 (\lnbar m_H^2 - 1/2) 
+ \fone (|\mu|^2, \stopone)
+ \fone (|\mu|^2, \stoptwo)
+ \fone (|\mu|^2, \sbotL) \Bigr ]/\sbeta^2
,
\eeq
where
\beq
\fone (x,y) &=& 
\left \{ \begin{array}{ll}
\left [2 x^2 \lnbar x + 2 (y - 2 x) y \lnbar y 
             + (x-y)(y-3x)\right ]/8 (x-y)^2
\phantom{xxxxx}
 & (x \not= y)\\[6pt]
(\lnbar x)/4 & (x=y),
\end{array}
\right.
\\
\ftwo(x,y) &=& 
\left \{ \begin{array}{ll}
 1 + y \ln(y/x)/(x-y)
\phantom{xxxxx}
 & (x \not= y)\\[-6pt]
\\
0 
 & (x = y).
\end{array}
\right.
\eeq
By comparing the one-loop gluon self-energy functions computed in both the
MSSM and the SM effective theory, and relying on the equality of physical
cross-sections computed in the two theories, one obtains: 
\beq
\cgA &=& \lnbar (m_{\tilde g}^2) -\frac{1}{2} 
       + \frac{1}{12} \sum_{i=1}^{12} \lnbar (m_{\tilde q_i}^2)
\eeq
Finally, by comparing the relevant two-loop part of $M_h^2$ in
eqs.~(\ref{eq:M2hMSSM})-(\ref{eq:M2htwoX}) above to the known result as
calculated directly in the MSSM \cite{EZ1,EZ2,SPM0206136,SPM0211366}, I
find: 
\beq
\cvev &=& 
-|\sstop \cstop|^2 (\stoptwo - \stopone)^2/(\stopone\stoptwo) 
+ |X_t|^2 \Bigl [
-3 (1 - 4 |\sstop \cstop|^2)
\stopone \stoptwo\ln(\stoptwo/\stopone)/2 (\stoptwo - \stopone) 
\nonumber \\ &&
+ 3 (\stopone + \stoptwo)/4
+ |\sstop \cstop|^2 (
  m^4_{\tilde t_1}/m^2_{\tilde t_2}
  + m^4_{\tilde t_2}/m^2_{\tilde t_1}
  - 7 \stopone - 7 \stoptwo)/2 
\Bigr ]/(\stoptwo - \stopone)^2 .
\label{eq:cv}
\eeq
\end{widetext}

The above results constitute a partial three-loop approximation to the
lightest Higgs mass in supersymmetry. A useful application of this, as has
been done earlier in \cite{Degrassi:2002fi}, is an estimate of the error
made in neglecting three-loop effects.  (See Appendix A for a comparison
of that paper and others with the results of the present paper.) However,
one would like to go further to use these results for an improved
calculation of the physical Higgs mass. To do so requires consistently
adding the three-loop correction to a more complete two-loop calculation
involving electroweak effects, which can be comparable in size. 

In earlier work, I have found the two-loop results for $M_h^2$ in the
MSSM, including all diagrammatic contributions to the pole mass that
involve the strong and Yukawa couplings (including those that also involve
electroweak couplings) \cite{SPM0405022}, as well as all of the remaining
contributions in the effective potential approach
\cite{SPM0206136,SPM0211366}. Since the results in the present paper are
also given in terms of running $\DRbarprime$ parameters, the three-loop
part can be consistently added to my previous results. However, there is
an important subtlety involving the identification of the tree-level Higgs
mass. In refs.~\cite{SPM0206136,SPM0211366,SPM0405022}, the tree-level $h$
squared mass is given by the appropriate eigenvalue of the $(H_u^0,H_d^0)$
squared mass matrix, evaluated with the VEVs at the minimum of the
two-loop effective potential. In the approximation used here, this
corresponds to: 
\beq
m_{h,{\rm tree}}^2 &=& m_h^2
- \frac{1}{2 v^2} \Bigl [ v_u \frac{\partial}{\partial v_u} + v_d \frac{\partial}{\partial v_d} \Bigr ] \Delta V_{\rm eff},
\phantom{xxx}
\label{eq:treeleveltad}
\eeq
where $m_h^2$ was defined by eq.~(\ref{eq:M2hzeroX}), and $\Delta V_{\rm
eff}$ is the radiative part of the effective potential, and the decoupling
approximation for the Higgs scalar bosons ($m_h^2 \ll m_{H^0}^2,
m_{H^\pm}^2, m_{A^0}^2$) has been used. The two versions of the tree-level
$h$ squared mass, $m_{h,{\rm tree}}^2$ and $m^2_h$, therefore differ by
tadpole loop contributions that involve the top Yukawa coupling and $g_3$. 

To avoid a mismatch between the two-loop part of the contribution found in
the present paper and the full two-loop contribution found in
refs.~\cite{SPM0206136,SPM0211366,SPM0405022}, one can take the results of
those papers and rewrite $m_{h,{\rm tree}}^2$ in the tree-level and
one-loop part in terms of $m_h^2$ as in eq.~(\ref{eq:treeleveltad}), and
then expand and incorporate the loop tadpole parts as residuals into the
one-loop and two-loop parts. In the two-loop part, one can consistently
simply replace $m_{h,{\rm tree}}^2$ by $m_h^2$. This is exactly what was
done above in the derivation of eq.~(\ref{eq:cv}), albeit in the
approximation of large $g_3, y_t$. (There are no technical obstacles to
this procedure in general, since the derivatives of the one-loop
self-energy functions with respect to the external momentum invariant and
the internal masses are well-known, and simple.) The resulting expression,
truncated at two-loop order, will then allow the three-loop contribution
of eq.~(\ref{eq:M2hthreeX}) to be added consistently. 

\section{Outlook} 
\setcounter{equation}{0} 
\setcounter{footnote}{1}

\begin{figure*}[t]
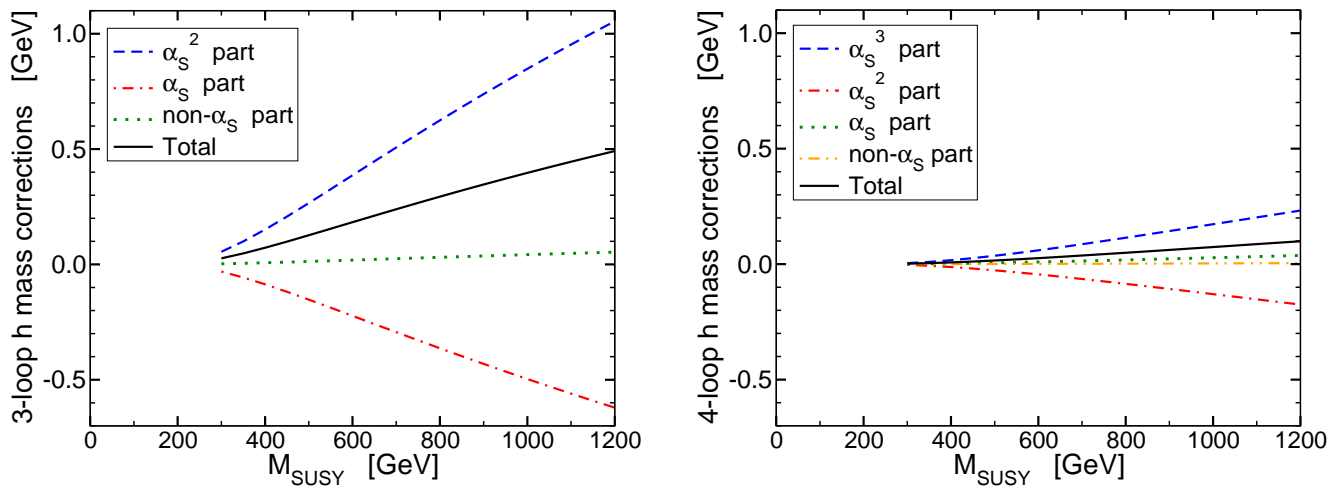

\centering
\mbox{\includegraphics[width=8.4cm]{delta3mh}}
\hspace{0.5cm}
\mbox{\includegraphics[width=8.4cm]{delta4mh}}
\caption{\label{fig:deltamh}
Leading-logarithm contributions to $M_h$ from 3 loops (left panel)
and 4 loops (right panel), as a function of the common superpartner
mass $M_{\rm SUSY}$. These results 
are due to 
the $L^3$ part of eq.~(\ref{eq:M2hthreeX}) 
and the $L^4$ part of eq.~(\ref{eq:M2hfourX}), 
respectively, with separate contributions from the various powers of
the MSSM QCD coupling and the total shown.
The Higgs pole mass is taken to be $M_h = 120$ GeV and $\tan\beta \gg 1$
and $\alpha_S^{(5),\MSbar}
(M_Z) = 0.118$ (in the five-quark effective Standard Model QCD theory) 
and top-quark pole mass $M_t = 172$ GeV.}
\end{figure*}

In this paper, I have evaluated the leading and next-to-leading logarithm
contributions to the lightest Higgs mass in the MSSM at three-loop order,
in the approximation of large QCD and top-quark Yukawa couplings. As
expected, these contributions are small, but still significant compared to
estimates of the experimental error for $M_h$ at the LHC or a future
linear collider. To show the size of the effects, I have plotted the
three-loop leading-log ($L^3$) contributions in the left panel of figure
\ref{fig:deltamh}, in the special limit of a common superpartner mass
$M_{\rm SUSY}$, and choosing $M_h = 120$ GeV and $\tan(\beta)\gg 1$, and
using $Q = M_{\rm SUSY}$ as the renormalization scale. [See
eq.~(\ref{eq:Mhsimp}).] The figure shows the separate contributions
proportional to $\alpha_S^2$, $\alpha_S$, and independent of $\alpha_S$,
as well as the total. There is a partial cancellation of the $\alpha_S^2$
and $\alpha_S$ parts, which is a fortuitous feature of the 
perturbative expansion scheme chosen here. 
This illustrates the more general fact that the numerical magnitude of the
three-loop correction depends on the way that perturbation theory is
organized. Changing the renormalization scale or scheme, or re-expanding
tree-level masses in one-loop and two-loop integral functions around pole
masses, can and does move contributions between loop orders.  For example,
ref.~\cite{Degrassi:2002fi} found the result quoted in
eq.~(\ref{eq:DHHSW}) of the present paper, which yields a rather larger
numerical magnitude for the three-loop correction, written in terms of
Standard Model effective couplings evaluated at $Q = M_t$. 

For comparison, the four-loop order leading logarithm ($L^4$)
contributions are shown in the right panel of figure \ref{fig:deltamh},
using the same scale. As expected, these are quite small, and there is
again a fortuitous partial cancellation between the leading and
next-to-leading orders in $\alpha_S$. 

The next-to-leading logarithm ($L^2$) contributions at three-loop order
can be seen to depend on the top-squark mixing and other details, and are
not depicted numerically here, although they can be significant. I expect
that in application to real-world data, one will want to re-express the
perturbative expansion by expanding tree-level masses in one-loop and
two-loop kinematic integrals around the pole masses, as this will likely
further improve the convergence of perturbation theory (see
ref.~\cite{Martin:2006ub} for an analogous discussion for the
gluino-squark system in supersymmetric QCD).  This will add more terms to
the next-to-leading parts of the three-loop contribution. 

If supersymmetry is discovered and thoroughly explored at the LHC and a
future linear collider, the mass of the lightest Higgs boson will present
an important precision test of our understanding of the theory. The
leading three-loop corrections will be important for this test. It should
be emphasized that, at this writing, some two-loop contributions remain
uncalculated, namely those involving purely electroweak couplings, which
may turn out to be similarly important. These contributions cannot be
captured adequately by the effective potential approximation, since some
of the relevant self-energy diagrams contain a routing by which the
external momentum does not go through any propagator with mass larger than
$M_Z$ or $M_h$. Therefore, it will probably be necessary to evaluate those
two-loop self-energy diagrams on-shell in order to reduce theoretical
errors to an acceptable level. 

\section*{Appendix: Comparison with other work} 

\renewcommand{\theequation}{A.\arabic{equation}}
\setcounter{equation}{0} 
\setcounter{footnote}{1}

It is useful to check the correspondence of the preceding formulas with
earlier special case results. For convenience, I focus on the two-loop
results given in refs.~\cite{EZ2} and \cite{ENHiggs}, and the three-loop
leading-log results of \cite{Degrassi:2002fi}. 

First, consider the approximations of refs.~\cite{EZ2}:
\beq
&&|\sstop \cstop| \>=\> 1/2;\\
&&m^2_{\tilde t_{1,2}} \>=\> M_{S}^2 \mp m_t |X_t|;\\
&&m_t^2,\> m_t |X_t| \>\ll\> M_S^2;\\
&&m_{\tilde g} \>=\> m_{\tilde b_L} \>=\> m_H \>=\> |\mu| = M_{S},
\eeq
but $|X_t|$ not necessarily small compared to $M_S$.
Then the results above become:
\beq
c_{y_t} &=& \frac{4}{3} \bigl 
[1 - {\rm Re}[X_t]/M_S + \lnbar (M_S^2)
\bigr ]
,
\\
c'_{y_t} &=& \frac{3}{4s_\beta^2} [(1 + c_\beta^2) \lnbar (M_S^2) - c_\beta^2/2]
,
\\
c_{\lambda} &=& 12 \lnbar (M_S^2) + 12 |X_t|^2/M_S^2 - |X_t|^4/M_S^4
,
\phantom{xxx}
\\
\cvev &=& |X_t|^2/4M_{\rm S}^2
,
\\
c_{g_3} &=& 2 \lnbar (M_S^2) - 1/2,
\eeq
in accord with the two-loop results found in eqs.(13)-(19) and (25) of
ref.~\cite{EZ2}. Note that the two-loop non-logarithmic parts in eqs. (16)
and (17) of ref.~\cite{EZ2} are not shown explicitly in the present paper; 
instead, the complete contributions up to two-loop order in
\cite{SPM0206136,SPM0211366,SPM0405022} can be included by the procedure
described at the end of section \ref{sec:matching}. 
 
Next, consider the approximations of ref.~\cite{ENHiggs},
which involve a light right-handed top squark:
\beq
&&m_{\tilde t}^2 \>=\> \begin{pmatrix} 
M_L^2 + m_t^2 & m_t X_t^* \\
m_t X_t & M_R^2 + m_t^2 
\end{pmatrix};
\\
&&m_t^2,\> m_t |X_t|\> \ll \> M_R^2 \>\ll \> M_L^2;
\\
&&m_{\tilde g} \>=\> m_{\tilde b_L} = m_H = |\mu| = M_{\rm L},
\eeq
but $|X_t|$ not necessarily small compared to $M_L$.
Then the results above become
\beq
c_{y_t} &=& \frac{1}{3} \bigl
[1 - 8 {\rm Re}[X_t]/M_L + 4 \lnbar (M_L^2)
\bigr ]
,
\\
c'_{y_t} &=& \frac{3}{8}
\Bigl (1 - \frac{2}{s_\beta^2}\Bigr ) [1 - 2 \lnbar (M_L^2)] 
- \frac{3 |X_t|^2}{4M_{\rm L}^2}
,
\\
c_{\lambda} &=& 6 \lnbar (M_L^2) + 6 \lnbar (M_R^2) 
+ \frac{12 |X_t|^2}{M_L^2}\ln(M_L^2/M_R^2) 
\nonumber \\ &&
+ \frac{6 |X_t|^4}{M_L^4} \bigl [2 - \ln(M_L^2/M_R^2) \bigr ]
,
\\
\cvev &=& 3 |X_t|^2/4M_{\rm L}^2,
\eeq
in agreement with the two-loop results of Appendix A and eqs. (B.8) and
(C.12) and (C.13) in ref.~\cite{ENHiggs}. Note that ref.~\cite{ENHiggs}
also explicitly identifies two-loop contributions enhanced by large
logarithms $\ln(M_L^2/M_R^2)$, which are not obtained in the approach used
here except when they are also enhanced by a $\ln(M_L^2/m_t^2)$. This is
because ref.~\cite{ENHiggs} used a multi-stage effective field theory
method, first decoupling left-handed top squarks and then right-handed top
squarks. Also, the approach of ref.~\cite{ENHiggs} isolates the
non-logarithmic two-loop effective potential contributions within the
given approximation, which are not shown explicitly here. Again, in the
approach of the present paper, the complete two-loop order results of
refs.~\cite{SPM0206136,SPM0211366,SPM0405022} can be included and
consistently supplemented by the three-loop result as described at the end
of section \ref{sec:matching}. 

Now, consider the three-loop leading log result given in eq.~(11) of
\cite{Degrassi:2002fi}, which assumes a single sparticle mass threshold at
$M_S$ and $\tan\beta \gg 1$, and is given in terms of running $\MSbar$
couplings evaluated at the top mass scale. In the notation of section
\ref{sec:SM} of the present paper, that result is: 
\beq
M_h^2 &=& \mZhat^2 +
\ythat^4 \vhat^2 \Bigl [
12 \kappa L_t 
+ \kappa^2 L_t^2 (- 96 \gthreehat^2 + 18 \ythat^2 ) 
\nonumber \\ &&
+ \kappa^3 L_t^3 (736 \gthreehat^4 - 240 \gthreehat^2 \ythat^2 - 99 \ythat^4)
\Bigr ],
\label{eq:DHHSW}
\eeq
where 
\beq
L_t = \ln(M_S^2/m_t^2) .
\eeq
Now, in the leading-logarithm approximation, the translation of SM
parameters evaluated at the top mass scale to the MSSM parameters
evaluated at $M_S$ depends only the Standard Model one-loop
renormalization group equations, and is independent of the $\MSbar$ to
$\DRbarprime$ conversion and threshold corrections. One can write: 
\beq
\gthreehat &=& g_3 \bigl [1 + \kappa L_t (7 g_3^2/2)  + ... \bigr ],
\\
\ythat &=& y_t \bigl [1 + \kappa L_t (4 g_3^2 - 9 y_t^2/4)
\nonumber \\ &&
           + \kappa^2 L_t^2 (22 g_3^4 - 18 g_3^2 y_t^2 + 243 y_t^4/32) + ...
           \bigr ],
\phantom{xxx}
\\
\vhat &=& v \bigl [1 + \kappa L_t (3 y_t^2/2) 
\nonumber \\ &&
           + \kappa^2 L_t^2 (6 g_3^2 y_t^2 - 9 y_t^4/4) + ...
           \bigr ],
\eeq
where the parameters on the left sides are evaluated at $Q = m_t $, and
those on the right sides at $Q = M_S$. Plugging these into
eq.~(\ref{eq:DHHSW}) immediately yields the special form of
eqs.~(\ref{eq:M2hMSSM})-(\ref{eq:M2hthreeX}) with only the leading
logarithms and $s_\beta = 1$: 
\beq
M_h^2 &=& m_Z^2 + y_t^4 v^2 \Bigl [
12 \kappa L_t 
+ \kappa^2 L_t^2 (96 g_3^2 -54 y_t^2 ) 
\nonumber \\ &&
+ \kappa^3 L_t^3 (736 g_3^4 - 672 g_3^2 y_t^2 + 90 y_t^4)
\Bigr ] .
\label{eq:Mhsimp}
\eeq
This is more useful as a rough indicator of the sizes of theoretical
errors due to three-loop effects than as an actual precise evaluation of
$M_h$. 

{\bf Acknowledgments:} This work was supported by the National 
Science Foundation under Grant No.~PHY-0456635.

\end{document}